\documentclass[journal]{IEEEtran}
\IEEEoverridecommandlockouts
\usepackage{amsmath,amssymb,amsfonts}
\usepackage{graphicx}
\usepackage{textcomp}
\usepackage{xcolor}
\usepackage{comment}
\usepackage{authblk}
\usepackage{multicol}
\usepackage[ruled,vlined,linesnumbered]{algorithm2e}
\usepackage{color}
\usepackage{graphicx}  % remove 'demo' option for your real document
\usepackage{soul}
\usepackage{physics}

\usepackage{comment}
\usepackage[numbers,sort&compress,square]{natbib}
\usepackage{enumerate}

\usepackage[numbers,sort&compress,square]{natbib}
\usepackage{pdfpages}

\usepackage[normalem]{ulem}
\usepackage{float}

\usepackage{scalerel}
\usepackage{tikz}
\usetikzlibrary{svg.path}

\definecolor{orcidlogocol}{HTML}{A6CE39}
\tikzset{
  orcidlogo/.pic={
    \fill[orcidlogocol] svg{M256,128c0,70.7-57.3,128-128,128C57.3,256,0,198.7,0,128C0,57.3,57.3,0,128,0C198.7,0,256,57.3,256,128z};
    \fill[white] svg{M86.3,186.2H70.9V79.1h15.4v48.4V186.2z}
                 svg{M108.9,79.1h41.6c39.6,0,57,28.3,57,53.6c0,27.5-21.5,53.6-56.8,53.6h-41.8V79.1z M124.3,172.4h24.5c34.9,0,42.9-26.5,42.9-39.7c0-21.5-13.7-39.7-43.7-39.7h-23.7V172.4z}
                 svg{M88.7,56.8c0,5.5-4.5,10.1-10.1,10.1c-5.6,0-10.1-4.6-10.1-10.1c0-5.6,4.5-10.1,10.1-10.1C84.2,46.7,88.7,51.3,88.7,56.8z};
  }
}

\newcommand\orcidicon[1]{\href{https://orcid.org/#1}{\mbox{\scalerel*{
\begin{tikzpicture}[yscale=-1,transform shape]
\pic{orcidlogo};
\end{tikzpicture}
}{|}}}}

\def\BibTeX{{\rm B\kern-.05em{\sc i\kern-.025em b}\kern-.08em
    T\kern-.1667em\lower.7ex\hbox{E}\kern-.125emX}}
\begin{document}

\title{Reduced Switching-Frequency Modulation Design for Model Predictive Control Based Modular Multilevel Converters}%*\\

%\title{Constrained Switching-Frequency Modulation for MPC-Based Modular Multilevel Converters}%*\\

%{\footnotesize \textsuperscript{*}Note: Sub-titles are not %captured in Xplore and
%should not be used}
%\thanks{Identify applicable funding agency here. If none, %delete this.}
%}

%\author{\IEEEauthorblockN{Saroj Khanal\orcidicon{0000-0002-0119-3188}\, \textit{Student Member, IEEE}, Vahid R. Disfani, \textit{Member, IEEE}}
\author{Saroj Khanal and Vahid R. Disfani \\
University of Tennessee at Chattanooga, Chattanooga, TN 37403 USA \\
emails: saroj-khanal@mocs.utc.edu, vahid-disfani@utc.edu 

\thanks{
\textcopyright 2019 IEEE. Personal use of this material is permitted. Permission from IEEE must be obtained for all other uses, in any current or future media, including reprinting/republishing this material for advertising or promotional purposes, creating new collective works, for resale or redistribution to servers or lists, or reuse of any copyrighted component of this work in other works. \\

Accepted and presented to the 2019 IEEE 2nd International Conference on Renewable Energy and Power Engineering.}}

%\thanks{Saroj Khanal and Vahid R. Disfani are with the ConnectSmart Research Laboratory and the Department of Electrical Engineering at the University of Tennessee at Chattanooga, Chattanooga, TN 37403 USA. {emails: saroj-khanal@mocs.utc.edu, vahid-disfani@utc.edu}}
%\author{Saroj Khanal \orcidicon{0000-0002-0119-3188}\,, \IEEEmembership{Student Member, IEEE}, Vahid R. Disfani \orcidicon{0000-0003-4727-6971}\,, \IEEEmembership{Member, IEEE}
%\author{Saroj Khanal, \IEEEmembership{Student Member, IEEE}, Vahid R. Disfani, \IEEEmembership{Member, IEEE}

%\thanks{Saroj Khanal and Vahid R. Disfani are with the Department of Electrical Engineering at the University of Tennessee at Chattanooga, Chattanooga, TN 37403 USA. {emails: saroj-khanal@mocs.utc.edu, vahid-disfani@utc.edu}}}
\maketitle

\begin{abstract}
This paper proposes a novel switching algorithm for modular multilevel converters (MMCs) that significantly reduces the switching frequency while fulfilling all control objectives required for their proper operation. Unlike in the conventional capacitor voltage-balancing strategies, in addition to submodule (SM) capacitor voltages, the proposed algorithm considers previous switching statuses during sorting. The algorithm is applied to a seven-level back-to-back MMC-HVDC system and tested under various operating conditions. Significant reduction in the switching frequency with trivial impacts on submodule capacitor voltages are observed. %\sarojreply{Vahid: is it fine to write numerical data in abstract? "Significant" might work in this case what we used in the first sentence.}
%\saroj{interesting things happened with {} with the block of abstract. It showed repeated extract with T removed in the beginging, and T added at the end -- interesting!: May be due to -- some things going on with class file!}
\end{abstract}

\begin{IEEEkeywords}
Capacitor voltage balancing, high voltage direct current (HVDC), model predictive control (MPC), switching algorithm.
\end{IEEEkeywords}

\section{Introduction}
Modular multilevel converter (MMC) stands out among converter topologies for medium- and high-power applications due to its salient characteristics such as modularity, scalability, high efficiency, high reliability, and improved power quality \cite{debnath2015operation, qin2012predictive, lesnicar}. Most popularly, it has become the worldwide standard for voltage-sourced converter high-voltage direct current (VSC-HVDC) transmission systems \cite{marquardt2018modular}.

Proper operation of MMC requires fulfillment of several control objectives, including control of output current, circulating current, and submodule (SM) capacitor voltage. Due to higher number of switching components, switching frequencies of MMCs are higher than those of VSCs, leading to undesirable power losses. Due to modularity, design of switching algorithms for MMCs is flexible and complex at the same time and is thence one of the most important technical challenges of MMC.

Several switching algorithms have been reported in the literature for MMC, including methods based on pulse width modulation (PWM) schemes and model predictive control \cite{debnath2015operation, ma2014one, disfani2015fast, du2017modular}. Among various switching methods, MPC-based methods have drawn significant attention as they offer fast dynamic performance with ability to meet multiple control objectives \cite{qin2012predictive,disfani2015fast, ma2014one, gong2016design, rodriguez2002multilevel, moon2014model}. Conventional MPC methods are computationally intensive and thus impractical, especially with high number of submodules (SMs). In addition, these MPC methods lead to high switching-frequency operation; thus, their adoption to leverage the benefits not only depends on reducing computational complexity but also on seeking reduced-switching frequency strategies \cite{dekka2017integrated}. In recent years, several computationally efficient MPC algorithms have been proposed to address the problem \cite{ma2014one, disfani2015fast,liu2015grouping,zhang2016voltage, huang2017priority}. However, little attention has been paid to reduce the switching frequency, which has a direct impact on converter loss and the reliability of switches.

In general, switching of MMC is performed in two stages: submodule sorting and submodule selection. In SM sorting stage, SMs are sorted based on their priority of being selected to be switched on in the next time step \cite{disfani2015fast, ma2014one, qin2012predictive}. Capacitor voltage balancing is the main objective which conventionally drives the sorting algorithm. In stage of SM selection, AC waveform tracking and circulating current mitigation are the objectives to be considered. This stage then determines how many of the sorted SMs should be selected to achieve them \cite{disfani2015fast, ma2014one, qin2012predictive}. One of the issues with these modulation techniques is their potentially high switching frequencies since the most sorting algorithms sort SMs just based on their voltage condition disregarding their current statuses (bypassed/inserted).

Several modulation strategies have been reported in the literature to reduce the switching frequency \cite{qin2013reduced, dekka2017integrated}.
%As one of the earliest efforts, a modified phase-shifted carrier-based PWM (PSC-PWM) together with a reduced-frequency algorithm is proposed in \cite{tu2011reduced},
%\vahid{the method in [18] is essentially doing what our S1-V2 algorithm does. Let's not cite it here in the Turkey paper. We will cite it in the other paper where we can talk about flexibility on $\overline{N_{sw}}$}\sarojreply{reply: I agree with you Vahid, we can remove. We are using F1-V2, instead of S1-V2, just in case you want to write it somewhere} \vahid{I made sure I used correct notations. Please remove [18]and clean the comments.}\sarojreply{Sure, I will do. Is there any way to bring the SM cap figure ahead of conclusions!, maybe at last it would work. I will try and inform you after I do it final}. which limits bypassing/insertion of the SMs every control cycle allowing turned-off SMs for next cycle to meet voltage requirement. \saroj{Merge this with last sentence at last: Though it considers current statuses of switches, its presentation is not scalable or readily available to be used for full-bridge MMC \vahid{We don't have FBMMC in this paper.} \sarojreply{reply: No, my point there was our algorithm is scalable than others. We can skip this if you think since were are not proving that.} and does not provide controllability over switching frequency and capacitor voltage compromise.}
A general framework for capacitor voltage balancing with reduced switching-frequency methods has been introduced in \cite{qin2013reduced}, where slow-rate, hybrid and fundamental-frequency capacitor-voltage balancing approaches are proposed to reduce switching frequency. However, the slow-rate method loses track of capacitor voltage. Dekka et al. in \cite{dekka2017integrated} propose an MPC platform to enable MMC to be modulated at a lower switching frequency; however, it does not explore further to reduce switching frequency.
%\vahid{can we provide more details why their method is no scalable/controllable?}. \sarojreply{Their way of presentation of results, I have having hard time to compare with ours. Their method and results are not focused on really evaluating switching frequencies and seeing its impact on SM Capacitor voltages. -- Can you look at their paper once? I will still look before you come to this point, will update if I am able to understand it.} \vahid{I can't get access to the paper. Do you have the pdf file of the paper to share with me?}\sarojreply{Just emailed you.}-- This has been repeated with what said before, merge both}

This paper investigates further opportunities to reduce switching frequency in a more effective and comprehensive way by proposing a novel modulation algorithm. The proposed algorithm takes the previous statuses of SMs into account while sorting the SMs for the selection process. It puts switched-on SMs in priority to be selected for the next time step, then sorts SMs with same current statuses based on their voltage values. Since voltage balancing is in lower priority in the proposed algorithm, capacitor voltage profiles of SMs might be compromised. %The more detailed analysis and description of the proposed algorithm can be found in \cite{khanal2019optimal}.
%The algorithm is designed in a way that it can be readily available to use for predictive methods and for MMCs with different SM topologies (e.g. full-bridge SMs based MMC) without any modifications. Further, the algorithm with with minor relaxation on the imposed constraints can provide the controllability over its trade-offs, which is suggested as a future work. 
This paper investigates the impacts of the proposed sorting algorithm on SM capacitor voltages to ensure the level of compromise is acceptable. Moreover, the comparison over conventional voltage-focused sorting algorithm is provided to demonstrate the effectiveness.

The rest of the paper is organized as follows. Section~\ref{sec:Math_model_MMC} details mathematical models of MMC. Section~\ref{sec:algorithms} explains the proposed novel reduced switching-frequency algorithm, which is tested against a standard MMC-based back-to-back HVDC system in Section~\ref{sec:case_study}. Finally, Section~\ref{sec:conclusion} concludes the paper.

\section{Mathematical Models for Modular Multilevel Converters} \label{sec:Math_model_MMC}
\subsection{MMC Topology}

\begin{figure}
\centering
\includegraphics[width=0.45\textwidth]{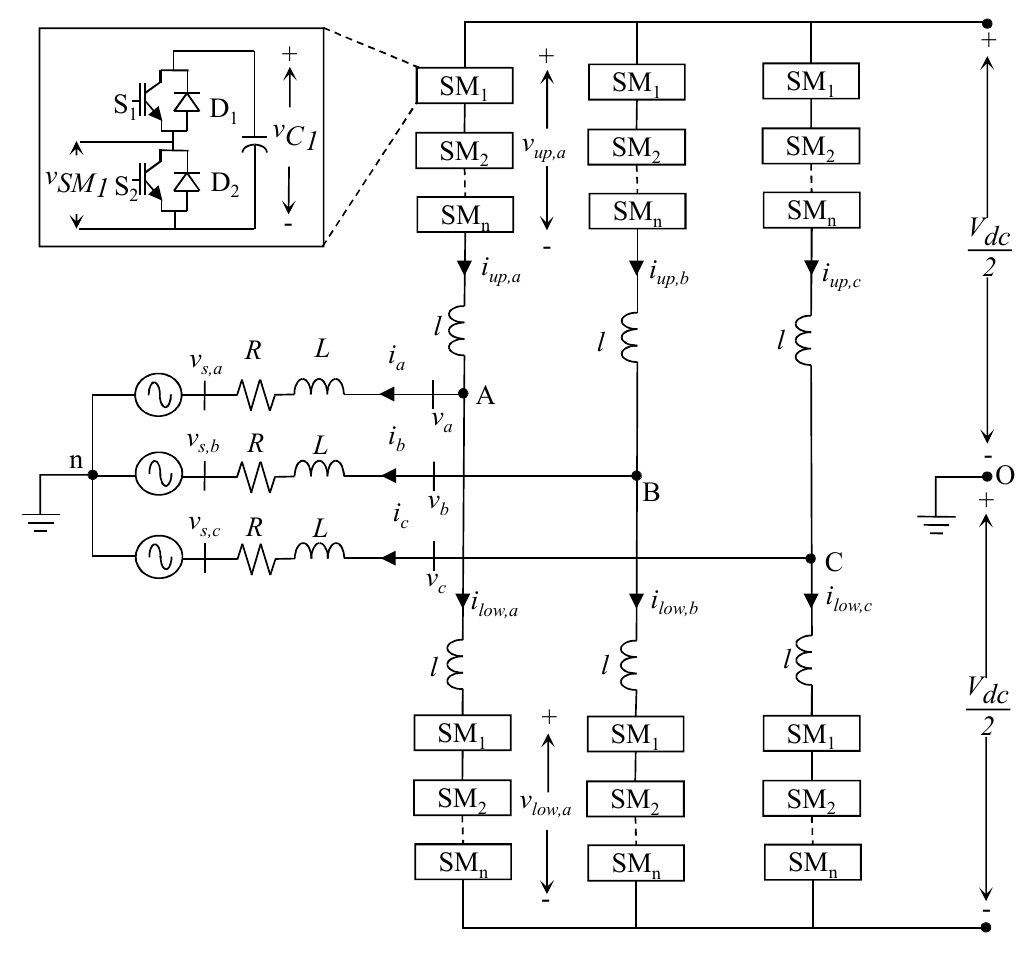}%MMC_config2}
\caption{Circuit diagram of a modular multilevel converter.}
\label{MMC}    
\end{figure}

Fig.~\ref{MMC} shows a schematic diagram of an $(n+1)$-level, three-phase MMC based on half-bridge SMs. Each phase (leg) consists of two arms, where each of them has of $n$ submodules (SMs). Among various SM configurations, half-bridge SM, which has two IGBT switches and a capacitor, is the most popular due to its simplicity and higher efficiency \cite{debnath2015operation}. MMC considered hereafter is MMC made up of half-bridge SMs. Arm inductors ($l$) are used for limiting current produced by instantaneous voltage difference and limiting fault currents. MMC is connected to a three-phase AC system through a series resistive-inductive ($R-L$) impedance. 

\subsection{Discrete Model of MMC} \label{MMC-discrete}
In this paper, the discrete model of MMC is built based on the authors' previous paper \cite{disfani2015fast}, where the next step value for the AC-side current for a sufficiently small sampling time step $T_s$ is derived as:
\begin{align}
&\begin{matrix}i(t+T_s)=\frac{1}{K'}{\left(  {\frac {v_{low}(t+T_s)-v_{up}(t+T_s)}{2} -v_s(t+T_s)+\frac{L'}{T_s}i(t)}\right)}\end{matrix}
\label{Idis}
\end{align}

\noindent where $L'=L+l/2$ and $K'=R+L'/T_s$. The measured values at the current time and the predicted values for the next time step are denoted by time indices $(t)$ and $(t+T_s)$, respectively. As the sampling frequency is assumed to be sufficiently higher than the grid frequency, the predicted value of grid voltage $v_s(t+T_s)$ can be replaced by its measured value $v_s(t)$. Defining $u_j(t+T_s)$ as the status of $j$-th SM, predicted capacitor voltage of individual SMs on upper-level and lower-level arms are equal to: 
\begin{align}
&v_{Cj}(t+T_s)=v_{Cj}(t)+\left(\frac{T_s i_{up}(t)}{C}\right)u_j(t+T_s) && \forall_{j\in [1,n]}\label{VCupdis} \\
&v_{Cj}(t+T_s)=v_{Cj}(t)+\left(\frac{T_s i_{low}(t)}{C}\right)u_j(t+T_s) &&\forall_{j\in [n+1,2n]}. \label{VClowdis}
\end{align}

Consequently, predicted voltages across upper-level and lower-level arms and circulating current are calculated as:
\begin{align}
&v_{up}(t+T_s)=\sum_{j=1}^{n} v_{Cj}(t+T_s)u_j(t+T_s)  \label{Vupdis} \\
&v_{low}(t+T_s)=\sum_{j=n+1}^{2n} v_{Cj}(t+T_s)u_j(t+T_s) \label{Vlowdis} \\
&i_z(t+T_s)=\frac{T_s}{2l}\left(V_{dc}-v_{low}(t+T_s)-v_{up}(t+T_s)\right)+i_z(t).
\label{Izdis}
\end{align}

\section{Reduced Switching-Frequency Model Predictive Control for MMC} \label{sec:algorithms}

\subsection{Optimization Model Formulation}
For an effective design of control systems and switching algorithms for MMC, the following four objectives are taken into account to \cite{qin2012predictive, disfani2015fast}:
\begin{enumerate}[i.]
\item regulate all the capacitor voltages on their nominal value ($V_{dc}/n$),
\item track the AC-side current ($i$) of all phases to their reference values ($i_{ref}$),
\item mitigate the circulating current ($i_z$) flowing among the converter legs, and
\item minimize the switching frequency.
\end{enumerate}

The first three objectives have been addressed in authors' previous works \cite{disfani2015fast,ma2014one} and is common among the predictive methods, while the contribution of this paper includes the addition of the last objective to the optimization problem.

%As discussed in \cite{disfani2015fast}, the constraint on the number of switched-on SMs increases the magnitude of circulating current significantly. In order to reduce these adverse impacts, this constraint is eliminated from the optimization problem~\eqref{originalopt}. %The modification enables the algorithm to switch on as many submodules as required to reach $v_{up}^*$ and $v_{low}^*$. 

With exact AC current waveform tracking $i(t+T_s)=i_{ref}$ and exact circulating current suppression $i_z(t+T_s)=0$, one could calculate the target values of upper-level and lower-level voltages of MMC as:
\begin{align}
&v_{up}^*=\left(\frac{V_{dc}}{2}+\frac{l}{T_s}i_z(t)\right)-
\left(K'i_{ref}+v_s(t)-\frac{L'}{T_s}i(t)\right)
\label{Vup*}\\
&v_{low}^*=\left(\frac{V_{dc}}{2}+\frac{l}{T_s}i_z(t)\right)+
\left(K'i_{ref}+v_s(t)-\frac{L'}{T_s}i(t)\right)
\label{Vlow*}
\end{align}
where $(\cdot)^*(t+T_s)$ denotes the ideal value of corresponding variable for the next time step. Defining $\Delta i=i-i_{ref}(t+T_s)$, $\Delta v_{low}=v_{low}^*-v_{low}$, and $\Delta v_{up}=v_{up}^*-v_{up}$, deviation of actual AC current waveform and circulating current from their target values are calculated as:

\begin{align}
&\Delta i= \frac{1}{2K'}\left(\Delta v_{low}(t+T_s)-\Delta v_{up}(t+T_s)\right)
\label{Idis_error}\\
&i_z(t+T_s)=\frac{T_s}{2l}\left(\Delta v_{low}(t+T_s)+\Delta v_{up}(t+T_s)\right)
\label{Izdis_error}
\end{align}

Applying weighted sum method to combine the second and third objective functions, the switching algorithm can be described as a multi-objective optimization problem with the formulation below:

\begin{align}
%&\min_U&& \left\{\begin{matrix}w_c{\Sigma_{j=1}^{2n}\abs{V_{C_j}(t+T_s)}}  + \\ \\
%w_u \Sigma_{j=1}^{2n} \abs{u_j(t+T_s)-u_j(t)} \end{matrix} \right\}  \label{sorting_obj}\\
&\min_U&& {\sum_{j=1}^{2n}\abs{v_{C_j}(t+T_s)-v_{C_j}(t)}}  \label{sorting_volt_obj}\\
&\min_U&& \sum_{j=1}^{2n} \abs{u_j(t+T_s)-u_j(t)}  \label{sorting_freq_obj}\\
&\min_U&&f=\left\{\begin{matrix}\frac{w}{2K'}\left|\Delta v_{low}(t+T_s)-\Delta v_{up}(t+T_s)\right|+\\
\\
\frac{w_z T_s}{2l}\left|\Delta v_{low}(t+T_s)+\Delta v_{up}(t+T_s)\right|\end{matrix}\right\}\label{selection_obj}\\
&\text{subject to:} &&~~\eqref{Idis}-\eqref{Izdis}\nonumber\\
%&&& ~~\sum_{j=1}^{n} \abs{u_j(t+T_s)-u_j(t)}\leq \overline{N_{sw}}\label{constraint_upper_sw}\\
%&&& \sum_{j=n+1}^{2n} \abs{u_j(t+T_s)-u_j(t)}\leq \overline{N_{sw}} \label{constraint_lower_sw}\\
%&&& ~~\underline{v_{C_j}} \leq v_{C_j}(t+T_s) \leq \overline{v_{C_j}}~~~~~~~~~~~~~~\forall_{j\in[1,2n]} \label{constraint_voltage_dev}\\
&&& ~~U=[u_1,u_2,...,u_{2n}] : u_j \in \{0,1\}~~~~\forall_{j\in[1,2n]}\label{constraint_status_binary}
\end{align}
where \eqref{sorting_volt_obj} regulates SM capacitor voltages, \eqref{sorting_freq_obj} minimizes number of switching events, and \eqref{selection_obj} follows the reference values of AC current and circulating currents. The values $w$ and $w_z$ are the weights applied to the objective functions corresponding to AC waveform tracking and circulating current elimination, respectively.

\subsection{MPC Solution Algorithms} 
\begin{figure}[H]
    \centering
    \includegraphics[width=0.45\textwidth]{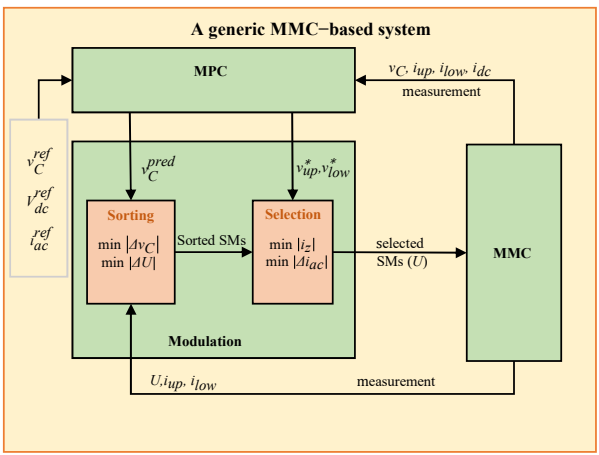}
    \caption{Control platform and information flow of an MPC-based modular multilevel converter.}
    \label{fig:MMCSys}
\end{figure}

Fig.~\ref{fig:MMCSys} illustrates control platform and information flow of an MPC-based modular multilevel converter. It includes two steps of SM sorting and SM selection, as detailed below, to solve the multi-objective optimization problem of the MPC algorithm. 

\subsubsection{Step 1 -- Submodule Cascaded Sorting} 
%\saroj{Vahid: Let's make this description custom to the algorithms we are presenting, especially at the last equations. We will be using this text quite often, maybe in the Australia paper, too. It be great if we customize this step by shortening it. Most of the things detailed in algorithms can be avoided wherever possible}
Since the MPC optimization problem must be solved in extremely short time steps, e.g. 100 $\mu s$, using regular optimization solution techniques such as interior point methods (IPM) and heuristic methods are not practical; the solution thus needs to be based on sorting algorithms. In this step, the objective functions \eqref{sorting_volt_obj} and \eqref{sorting_freq_obj} are targeted by sorting SMs effectively such that the highest priority is given to the SMs contributing the most in voltage balancing and switching frequency reduction. 

To address \eqref{sorting_volt_obj}, SMs of both upper and lower arms are sorted based on their anticipated capacitor voltages. According to \eqref{VCupdis}, the direction of $i_{up}$ defines whether the capacitor voltages of upper-level submodules are subject to increase or decrease. Therefore, the corresponding algorithm sorts SMs based on their capacitor voltages in ascending order if $i_{up}\ge 0$, and in descending order if $i_{up}<0$. On the other hand, to minimize \eqref{sorting_freq_obj}, the switching algorithm sorts SMs based on their current status $U(t)$. To minimize the switching frequency, the algorithm gives high priority to the SMs that are currently switched on and low priority to the ones currently switched off.

This procedure is detailed in \textbf{Algorithm~\ref{F1-V2}} and is called F1-V2 hereafter. Conventional sorting algorithm focused on voltage balancing is called V1-F2 in this paper and is used as the benchmark algorithm in the case study.

\begin{comment}

\begin{algorithm}%[H]
\caption{Voltage Balancing Algorithm (V1-F2)}
\label{V1-F2}
\begin{algorithmic}
\FOR{all phases $a,b,c$}
\STATE {Collect measurements of capacitor voltages, arm currents, and current switching status ($u_j^{curr}$).}
\STATE Calculate anticipated SM voltages $v_{C_j}$
\FOR{$k\in\{up,low\}$}
\STATE Sort SMs based on $u_j^{curr}$ in descending order
\IF{$i_{k}>0$}
\STATE Sort sorted SMs based on $v_{C_j}$ in descending order
\ELSE 
\STATE Sort sorted SMs based on $v_{C_j}$ in ascending order
\ENDIF
\ENDFOR
\ENDFOR
\STATE Proceed to \textbf{Step 2}
\end{algorithmic}
\end{algorithm}

\begin{algorithm}%[H]
\caption{Reduced-Switching-Frequency Algorithm (F1-V2)}
\label{F1-V2}
\begin{algorithmic}
\FOR{all phases $a,b,c$}
\STATE {Collect measurements of capacitor voltages, arm currents, and current switching status ($u_j^{curr}$).}
\STATE Calculate anticipated SM voltages $v_{C_j}$
\FOR{$k\in\{up,low\}$}
\IF{$i_{k}>0$}
\STATE Sort SMs based on $v_{C_j}$ in descending order
\ELSE 
\STATE Sort SMs based on $v_{C_j}$ in ascending order
\ENDIF
\STATE Sort sorted SMs based on $u_j^{curr}$ in descending order
\ENDFOR
\ENDFOR
\STATE Proceed to \textbf{Step 2}
\end{algorithmic}
\end{algorithm}
\end{comment}

\begin{algorithm}%[H]
\DontPrintSemicolon
\caption{Reduced-Switching-Frequency Voltage-Balancing Algorithm (F1-V2)}
\label{F1-V2}
\For{all phases $a,b,c$}{
    Collect measurements of capacitor voltages, arm currents, and current switching status ($u_j^{curr}$)\;
    Calculate anticipated SM voltages $v_{C_j}$\;
        \For{$k\in\{up,low\}$}{
        \eIf{$i_{k}\ge 0$}
        {Sort SMs based on $v_{C_j}$ in ascending order\;}
        {Sort SMs based on $v_{C_j}$ in descending order\;}}
    Sort the sorted SMs based on $u_j^{curr}$ in descending order\;}
Proceed to \textbf{Step 2}.
\end{algorithm}
 %Also, let function $G$ map the index of each SM to its order in the sorted vector $[V_{C_{up}}^{sort},V_{C_{low}}^{sort}]$.
\subsubsection{Step 2 -- Submodule Selection}
Submodule selection algorithm is formulated mainly based on authors' previous paper \cite{disfani2015fast}. Let the vectors $V_{C_{up}}^{sort}=[V_{C_1}^{sort},...,V_{C_n}^{sort}]$ and $V_{C_{low}}^{sort}=[V_{C_{n+1}}^{sort},...,V_{C_{2n}}^{sort}]$ denote SM voltages on upper and lower arms, respectively, after being sorted in Step 1. In this step, the algorithm first calculates the cumulative sum vectors of the components of $V_{C_{up}}^{sort}$ and $V_{C_{low}}^{sort}$. The sets of cumulative sum values are denoted as $V_{C_{up}}^{sum}$ and $V_{C_{low}}^{sum}$, and are defined as below.
\begin{align}
&V_{C_{up}}^{sum}=\{\alpha_k:k=0,1,...,n\} \label{sum_up}\\
&V_{C_{low}}^{sum}=\{\beta_k: k=0,1,...,n\}  \label{sum_low}
\end{align}
where
\begin{align}
&\alpha_0=\beta_0=0\nonumber\\
&\alpha_k=\Sigma_{i=1}^{k}V_{C_i}^{sort}&\forall_{k\in [1,n]}\nonumber\\
&\beta_k=\Sigma_{i=n+1}^{n+k}V_{C_i}^{sort}&\forall_{k\in [1,n]}\nonumber
\end{align}
The switching algorithm then defines what combination of $(\alpha,\beta)$ minimizes the objective function \eqref{selection_obj}. It has been proven in \cite{disfani2015fast} that if $v_{up}^*\in [\alpha_i,\alpha_{i+1})$ and $v_{low}^*\in [\beta_j,\beta_{j+1})$, the optimal solution belongs to the set $\{(\alpha_i,\beta_j),(\alpha_{i+1},\beta_j), (\alpha_i,\beta_{j+1}),(\alpha_{i+1},\beta_{j+1})\}$. It means that it suffices to check the objective function for just 4 points instead of $n^2$ feasible solutions to select the best SMs to switch on.

%\subsection{Reliability-Aware Switching Algorithm (F1-V2)}

%For the reliability consideration in switching, we have proposed the algorithm shown in \textbf{Algorithm \ref{V1-F2}}. 

\begin{comment}

\begin{algorithm}[H]
\caption{Step 2: SM selection}
\begin{algorithmic}
\STATE Create the vectors $V_{C_{up}}^{sort}$ and $V_{C_{low}}^{sort}$ based on the output of \textbf{Algorithm~\ref{V1-F2}} or \textbf{Algorithm~\ref{F1-V2}}.%\eqref
\STATE Define $i$ such that $\alpha_i\leq v^*_{up}\leq\alpha_{i+1}$
\STATE Define $j$ such that $\beta_j\leq v^*_{low}\leq\beta_{j+1}$
\STATE Let $k=0$
\FOR{any $(\alpha,\beta)\in\{\alpha_i,\alpha_{i+1}\}\times\{\beta_j,\beta_{j+1}\}$}
\STATE Calculate $A_k=f(O(\alpha,\beta))$ using \eqref{selection_obj}
\STATE Let $k=k+1$.
\ENDFOR
\STATE Find the minimum value of $A_k$ and report the corresponding sequence as the final solution $U^*$.
\end{algorithmic}
\end{algorithm}

\end{comment}

\section{Case Study} \label{sec:case_study}
\subsection{Test System}
\begin{figure}[H]
    \centering
    \includegraphics[width=0.45\textwidth]{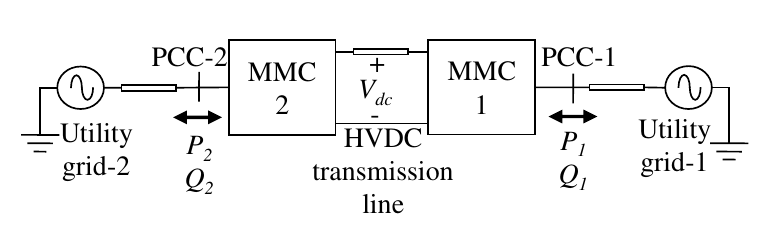}
    \caption{Schematic representation of an MMC-based back-to-back HVDC System.}
    \label{fig:HVDC_SLD}
\end{figure}

In this section, the proposed reduced switching-frequency algorithm is applied against a back-to-back MMC-based HVDC system, as shown in Fig.~\ref{fig:HVDC_SLD}. The test system parameters are provided in Table~\ref{tab:sys_param}. The simulation is run for 3 seconds, starting with the V1-F2 switching algorithm. At $t=1.2~\mathrm{s}$, the switching algorithm is changed to F1-V2; at $t=1.4~\mathrm{s}$, it is changed back to the V1-F2 algorithm. The effectiveness of the algorithm is benchmarked against the standard conventional voltage-centered sorting method (V1-F2).

\begin{table}[H]
    \centering
    \caption{Case study parameters}
    \begin{tabular}{|l|l|}
    \hline
        Quantity & Value \\
        \hline
        Number of submodules per arm & 6 \\ 
        MMC nominal power & 50 MVA \\
        Nominal DC voltage $(V_{dc})$ & 60 kV \\
        Submodule capacitor ($C_{sm}$) & 2.5 mF\\
        %Carrier signal frequency ($f$) & 20 kHz\\
        Active power transferred ($P_{1}$) & 13.18 MW \\
        $R$ & 0.03 $\Omega$\\
        $L$ & 5 mH\\
        $l$ & 3 mH\\
        Sampling period ($T_s$) &25 $\mu$s\\
        HVDC line length &5 km\\
        HVDC link capacitor &16 $\mu$F/km\\
        HVDC line inductance &50 $\mu$H/km\\
        \hline
    \end{tabular}
    \label{tab:sys_param}
\end{table}

\subsection{Results and Discussions}

%Table~\ref{tab:fs_table} shows steady-state average effective switching frequencies ($f_s$) observed with both switching algorithms as in Fig.~\ref{fig:f_s}. It clearly shows 83\% effective $f_s$reduction in upper-arm, first SM and 78\% in lower-arm, first SM of phase A.

%\saroj{This table is not required anymore. Should be removed!}
\begin{comment}

\begin{table}
    \centering
    \caption{Effective switching frequency ($f_S$) with various algorithms}
    \begin{tabular}{|l|l|p{2.125cm}|p{2.125cm}|}
    \hline
    Time & Algorithm & $f_s$ of first upper-arm SM of phase A & $f_s$ of first lower-arm SM of phase A \\
    \hline
     \hline
    [8.8 s, 9.0 s] & V1-F2 & 6755 Hz & 6742 Hz \\
    \hline
    [9.02 s, 9.18 s] & F1-V2 & 1149 Hz & 1316 Hz \\
    \hline
    [9.22 s, 9.4 s] & V1-F2 & 6659 Hz & 6818 Hz \\ 
    \hline
    \end{tabular}
    \label{tab:fs_table}
\end{table}
\end{comment}

\begin{figure}
    \centering
    \includegraphics[width=0.48\textwidth]{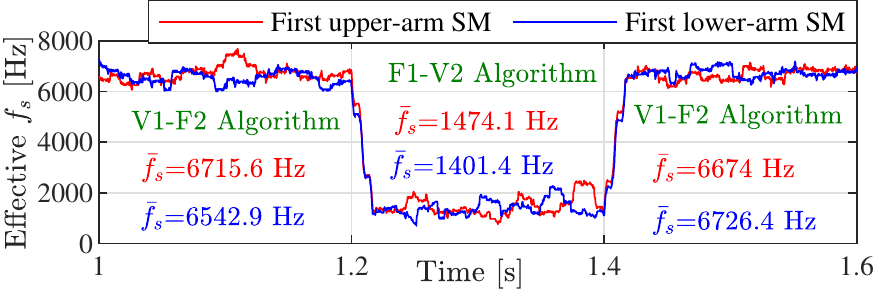}
    \caption{Effective switching frequencies of selected SMs of upper- and lower-arms using both V1-F2 and F1-V2 algorithms.}
    \label{fig:f_s}
\end{figure}

Fig.~\ref{fig:f_s} shows the effective switching frequencies while applying both the algorithms. With V1-F2 algorithm, the results demonstrate that the effective switching frequency ($f_s$) is equal to 6715.6~Hz and, then, it drops to 1474.1~Hz with F1-V2 algorithm. That is, the switching frequency ($f_s$) in F1-V2 is about 22\% of that in V1-F2, resulting in 78\% reduction in switching losses and 390\% improvement in lifetime of switches.

\begin{figure*}
   \centering
   \includegraphics[width=0.8\textwidth]{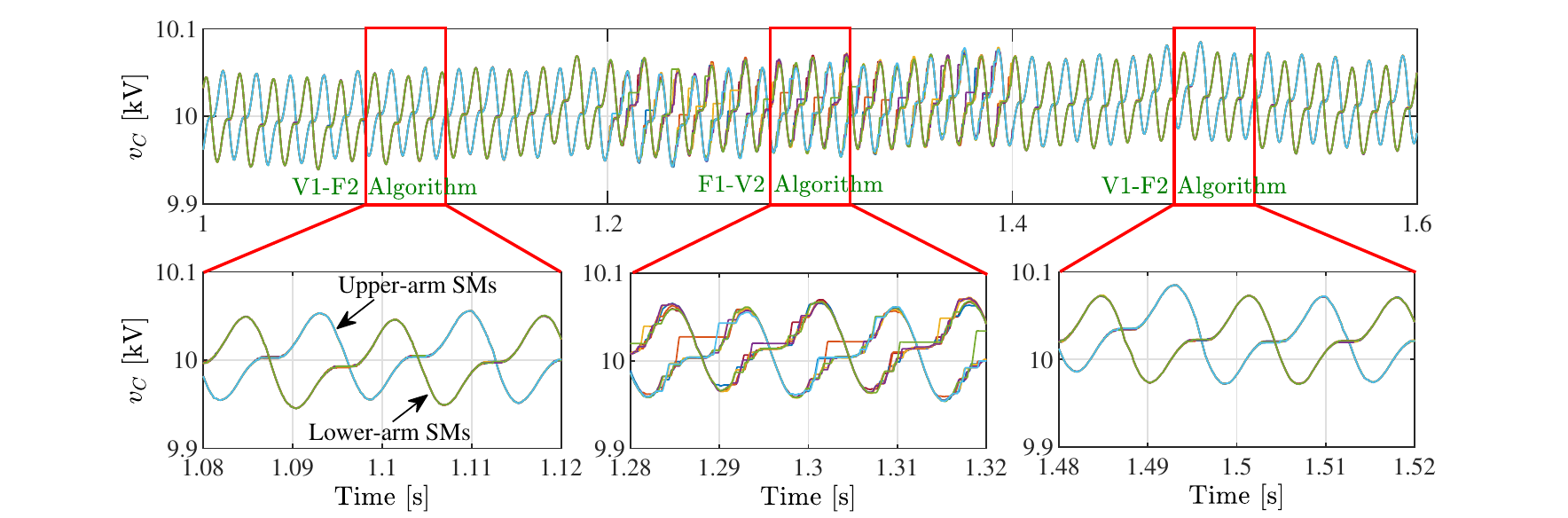}
   \caption{SM capacitor voltages of lower and upper arms of phase A of MMC1 for both V1-F2 and F1-V2 algorithms.}
   \label{fig:v_C}
\end{figure*}

Fig.~\ref{fig:v_C} shows the capacitor voltages of both the upper and the lower arms of the phase of MMC1. The switching algorithm has some effects on how these voltage curves behave. With V1-F2 algorithm, the capacitor voltages of all submodules on the upper (or the lower) arm change altogether as a result of the voltage balancing strategy employed in this paper. Between $t=1.2~\mathrm{s}$ and $t=1.4~\mathrm{s}$, there are some differences between individual capacitor voltages of submodules of the upper (or lower) arm. This is because the priority in this switching algorithm is given to reduce the switching frequency of the MMC, and thence the voltage balancing objective is compromised. The main takeaway is that the voltage ripples of SM capacitors are all the same (equal to 1.2\%) and are not affected by the switching algorithm employed.

\begin{figure}[H]
     \centering
     \includegraphics[width=0.48\textwidth]{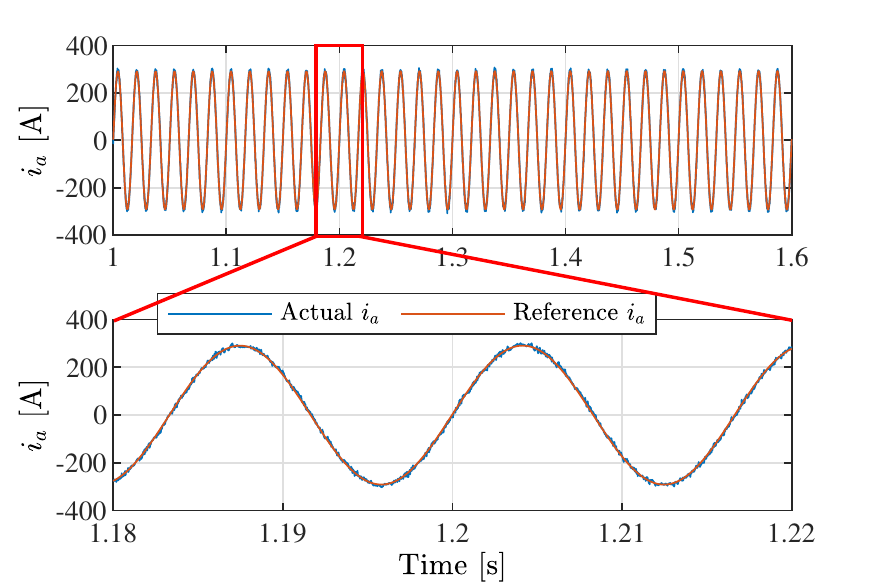}
     \caption{AC terminal current at phase A of MMC1.}
     \label{fig:I_a}
\end{figure}

Fig.~\ref{fig:I_a} depicts the reference and actual output AC current at the terminal A of MMC1. The results show that the AC current perfectly follows the reference waveform, fulfilling the second objective for both switching algorithms.

\begin{figure}[H]
    \centering
    \includegraphics[width=0.48\textwidth]{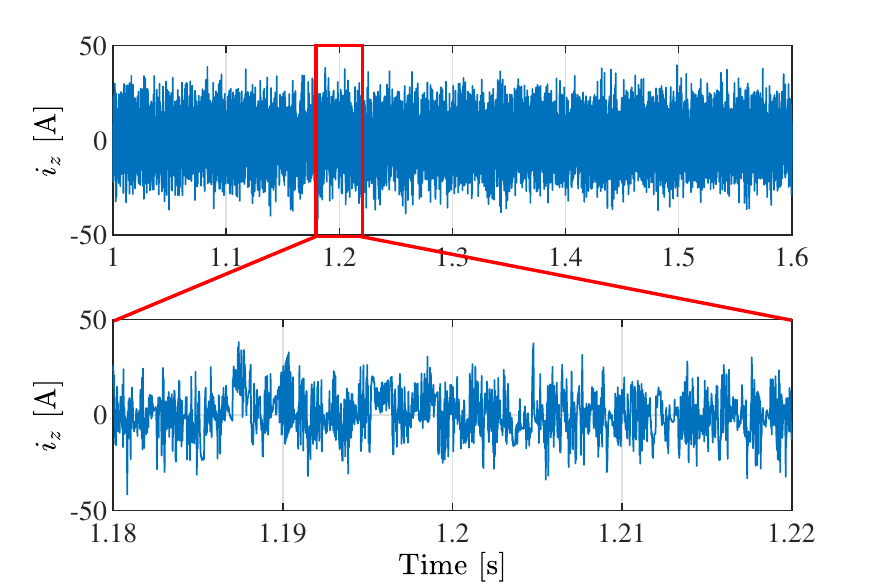}
    \caption{Circulating current of MMC1.}
    \label{fig:I_z}
\end{figure}

Fig.~\ref{fig:I_z} illustrates the circulating current of phase leg A of MMC1 at [1 s, 1.6 s] of simulation. The results show that the circulating current is successfully controlled around zero Ampere, and its maximum deviation from zero is just 10\% of the magnitude of output AC current of MMC1. These results depict how the third objective function of mitigating circulating current is fulfilled for both switching algorithms. Zoomed view of the circulating current is also shown in Fig.~\ref{fig:I_z}. It represents a transitional time from V1-F2 algorithm to F1-V2 algorithm. There is no negative effect of the proposed algorithm on the circulating current.

\begin{comment}

Fig.~\ref{fig:P1} depicts active power transferred by MMC1 to the utility grid-1. It shows the active power delivery of 13.45 MW.
\begin{figure}[H]
    \centering
    \includegraphics[width=0.48\textwidth]{figures/P_1.pdf}
    \caption{The output active power of MMC1 to the utility grid-1.}
    \label{fig:P1} 
\end{figure}
\end{comment}

\begin{figure}[H]
    \centering
    \includegraphics[width=0.45\textwidth]{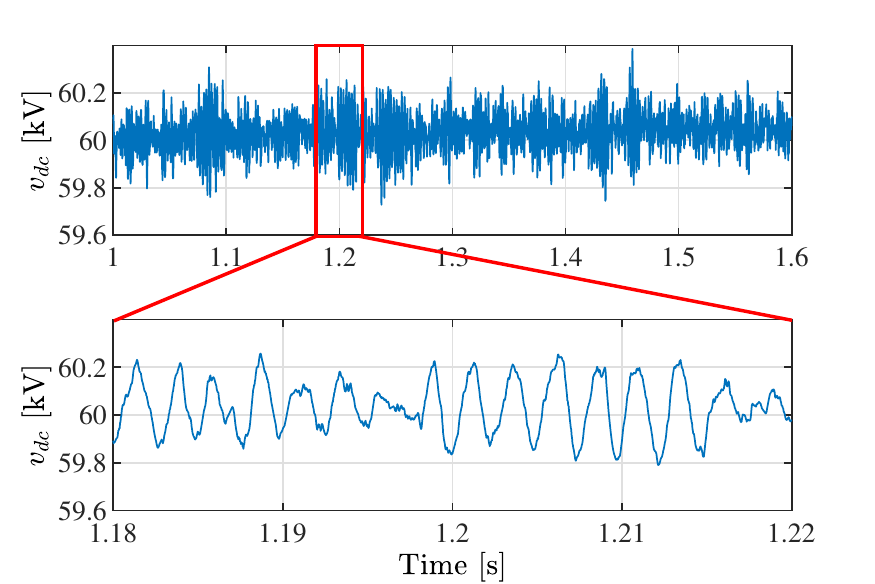}
    \caption{HVDC link voltage.}
    \label{fig:V_dc}
\end{figure}

\begin{figure}[H]
    \centering
    \includegraphics[width=0.45\textwidth]{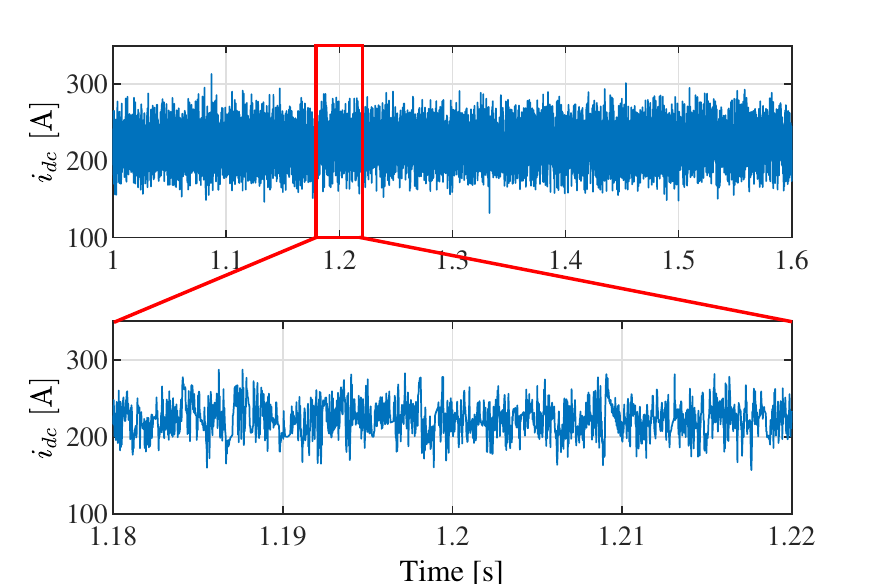}
    \caption{HVDC link current from MMC2 toward MMC1.}
    \label{fig:I_dc}
\end{figure}

Fig.~\ref{fig:V_dc} and Fig.~\ref{fig:I_dc} show the DC link voltage and the DC link current, respectively. Results demonstrate that the voltage and the current are satisfactorily around their nominal values of 60~kV and 225~A, respectively, and are not compromised because of the proposed algorithm.

\section{Conclusion} \label{sec:conclusion}
This paper proposes a novel modulation technique to reduce the switching frequency of submodules (SMs), where the sorting algorithm prioritizes SMs based on their current statuses. SM selection algorithm is based on a model predictive control (MPC) platform with the multiple control objectives on SM capacitor voltage balancing, AC output current tracking, and circulating current mitigation. The proposed algorithm is tested against a three-phase MMC-based back-to-back HVDC system in MATLAB/Simulink and is benchmarked against the conventional voltage-balancing algorithm. The proposed algorithm reduces the switching frequency by 78\% while satisfying all MPC control objectives, with trivial impacts on SM capacitor voltages.

\bibliographystyle{ieeetr}
{ \footnotesize \bibliography{IEEEabrv,MMC}}

\end{document}